\documentclass[12pt]{article}
\usepackage{amsmath,amssymb}
\usepackage{graphicx}
\usepackage{caption}
\usepackage{color}
\usepackage{dcolumn}
\usepackage{bm}
\usepackage[numbers,super,comma,sort&compress]{natbib}

\usepackage{xcolor}
\usepackage{float}
\usepackage{tikz}
\usepackage{pgfplots}

\def\br{\begin{eqnarray}}
\def\er{\end{eqnarray}}
\def\be{\begin{equation}}
\def\ee{\end{equation}}
\def\({\left(}
\def\){\right)}

\def\ii {\'\i  }

\title{Study of Quantum Confinement inside a Viral Capsid}
\author{Elso Drigo Filho\thanks{Instituto de Bioci\^encias, Letras e Ci\^encias
Exatas, IBILCE-UNESP,  S\~ao Jos\'e do Rio Preto, SP, Brazil}, Regina Maria Ricotta \thanks{Faculdade de Tecnologia de S\~ao Paulo, Fatec-SP-CEETPS-UNESP, S\~ao Paulo, SP, Brazil}}
\begin{document}
\maketitle
\begin{abstract}

Classical computational methods, such as molecular dynamics and Monte Carlo simulations, have long been the standard for modeling viral structure and function. However, these approaches may overlook crucial quantum phenomena that operate at the nanoscale, particularly within the highly-compacted genetic material of the viral capsid. The confined, high-density environment of genetic material within the capsid strongly suggests that quantum confinement effects play a significant, yet unexplored, role in viral processes. This study introduces a novel quantum approach using Supersymmetric Quantum Mechanics (SQM) to investigate the quantum confinement effects on viruses. In this paper, the viral capsid environment is modeled using the Pariacoto virus, a model system well-suited for this analysis due to its specific structural properties. The findings reveal that quantum effects are not merely marginal but essential for understanding key processes inside the capsid, providing new insights beyond the scope of classical physics.\\
Keywords: Viral Capsid, Supersymmetric Quantum Mechanics, Quantum Confinement, Debye-H\"{u}ckel Potential, Variational Method.

\end{abstract}

  \makeatletter
  \renewcommand\@biblabel[1]{#1.}
  \makeatother

\bibliographystyle{apsrev}

\renewcommand{\baselinestretch}{1.5}
\normalsize

\section{Introduction}
Viruses are structures formed by genetic material (nucleic acid) surrounded by a protein shell called a capsid and, occasionally, by an outer lipid envelope.  Although theoretical treatments involving viral properties have been conducted, most results stem from computer simulations \cite{Robson, Shi, Perlmutter} based on essentially classical tools. 

The general field of quantum biology, which explores the potential for non-trivial quantum effects in biological systems, is an area of growing interest \cite{McFaddenJ}. The inclusion of quantum effects in virology  has been considered in specific contexts, such as in the analysis of quantum tunnelling and quantum dots with applications in virology \cite{Adams, Roy}. It is crucial to note that neglecting quantum contributions has already been identified as a potential source of error in simulations, as indicated by Robson  \cite{Robson}. Severe spatial confinement (strong confinement) can introduce distortions in the quantum eigenstates, which are important for describing the phenomena occurring inside the viral capsid.

Quantum confinement can be induced by restrictions on the molecular structure of the receptacle surrounding a molecule, as in the case of nanotubes  \cite{Wang, Kim} or zeolites  \cite{Wang, Stucky}.  Anomalous effects can also be observed in systems under high pressure, a factor that may have played a significant role in the origin of life \cite{Bassez, Daniel}. Studies related to osmotic pressure indicate that the osmotic pressure within a viral capsid can reach values of up to 100 atmospheres \cite{Smith1, Cordova, Siber}.

In biological systems, polyelectrolytes are abundant. The electrostatic shielding effect is fundamental for the structure and function of these molecules in living organisms. In particular, nucleic acids  \cite{Manning} and proteins \cite{Warshel}  depend on electrostatic properties to function within their biological environment.

In this regard, we identify at least two important effects acting simultaneously inside the viral capsid: the abundance of electrical charges and components under high pressure \cite{Khaykelson, Pollard}. Therefore, to understand the molecular behavior within the viral capsid, it is essential to investigate the interplay between these factors and establish the relevance of confinement effects for these systems.

Using a simplified model for the internal environment of a virus and a typical value for the osmotic pressure, it is possible to estimate the available molecular volume for each component. From this volume, we can calculate the relevance of the quantum confinement effect. A widely used potential to describe the screened Coulomb interaction in biological molecules is the Debye-H\"{u}ckel potential, \cite{Shi,Perlmutter,Forrey,Perego}. The Pariacoto virus is used to illustrate this approach \cite{Forrey}.

The quantum effects description in the internal material of the viral capsid has not been considered in the models used so far, \cite{Robson,Siber,Martinez}. This omission is understandable when considering the total size of the capsid (distances on the order of tens of nanometers). Forrey and Muthukumar \cite{Forrey}, for instance, use the distance from the center to the face of the Pariacoto virus capsid as $13.8\, nm$. This approach fails to account for the fact that the material inside the virus is subject to high pressure due to the high density of particles. This leads us to the central hypothesis of this work: quantum confinement effects introduce distortions in the quantum eigenstates and are important for describing these phenomena.

In this paper, the effective confinement volume within the viral capsid is estimated using the ideal gas model as a first approximation (section 2). Subsequently, the variational method associated with the algebraic method of supersymmetric quantum mechanics is used to determine the energy eigenvalue of the ground state of the Debye-H\"{u}ckel potential for different confinement radii (section 3). The parameters of the Pariacoto virus are used in the description of the viral capsid; the results are presented in section 4. Finally, section 5 contains the conclusions.

\section{Molecular volume within the capsid} 

In the proposed model, the internal environment of the virus consists of particles that behave like an ideal gas. This approach is discussed at length by Zuckerman \cite{Zuckerman}. This simplification allows us to obtain an analytical expression for the molecular volume directly. Crucially, the inclusion of interactions between particles within the capsid should not invalidate the conclusions reached here regarding the need to incorporate quantum confinement effects in describing this environment. Rather, both internal interactions (e.g., dipole interactions) and the physical dimensions of the molecules (such as molecular radius and hydration layers) further intensify the internal conditions of the virus. In this way, these conditions reinforce the primary conclusion of this study. 

For modeling purposes, water is considered an inert environment, a common practice in computer simulations using mean-field theory. In these cases, water contributes only a minor value to the overall dielectric constant. The osmotic pressure difference between the highly dilute external environment and the virus's usual internal system is approximately $100\,atm = 1.0\,.10^7\, Pa$ \cite{Smith1, Cordova, Siber}. This value is adopted as a typical pressure inside the viral capsid.

We model the molecules in solution as an ideal gas. In this case,  the ideal gas law, $PV = nRT$, applies, where $P$ is the pressure (adopted here as being $100 \,atm$), $V$ is the volume, $n$ is the number of particles, $R = 0.082 atm.l/mol.K$ is the ideal gas constant and $T$ is the temperature (assumed to be $T = 300 K$). Then, we get
\br
\label{particles}
\frac{n}{V}=\frac{P}{RT} = 4.065\, \frac{mol}{l} = 2.447 \;.10^{24}\, \frac{particles}{l}= 2.447\,.10^{-3}\, \frac{particles}{{\AA}^3}
\er

Under these conditions, we can estimate that each particle occupies a volume of about $409 \AA^3$ or $2747 a_0^3$, where $a_0$ is the Bohr radius ($a_0 = 0.53 \AA$).

\section{The Schr\"{o}dinger equation for the confined Debye-H\"{u}ckel potential}

Assuming a central force interaction, each particle of volume $409 \AA^3$   would be confined within a spherical box with a radius of approximately $R_C=4.6 \AA$ (or $8.7 a_0$). Previous analyses using a Morse potential to simulate molecular vibration have shown that a confinement distance of this magnitude is sufficient to induce observable quantum effects in the ground state, as demonstrated for both simple hydrogen \cite{Silva1} and more complex diatomic molecules \cite{Drigo1}. However, for longer-range interactions, such as those described by the Debye-H\"{u}ckel potential, a more careful analysis is required.

	The Debye-H\"{u}ckel potential can be written in the following form \cite{Forrey}:
\br
\label{DH-potential}
V(r) = \frac{\delta}{r} e^{-\kappa r}
\er
where $\delta = l_B k_B T = 1.2 \,. 10^{-2}\, u.a.$ (atomic units), $\kappa^{-1}= 0.7 nm = 13 a_0 $ is the Debye length and $a_0$ is the Bohr radius, $l_B = 0.7\, nm = 13 a_0$ is the length of Bjerrum, $k_B T = 4.1 \,pN \,nm = 9.5 \,. 10^{-4} \,a.u.$. These data come from the reference\cite{Forrey} and are used in the analysis of the Pariacoto virus.

In atomic units, ${\hbar}=m=e=1.0$, the Schr\"{o}dinger equation is written as follows:
\br
\label{SE}
\hat{H}\Psi (r) =(-\frac{1}{2} \frac{d^2}{ dr^2} +V(r) )\Psi (r) = E \Psi (r)
\er
where  $\hat{H}$  is the Hamiltonian operator, $\Psi(r)$ is the wave function of each particle within the viral capsid and E corresponds to the energy eigenvalues. For the purposes of the present paper only the ground state energy is analyzed. 

The problem to be solved is to obtain the solution of equation (\ref{SE}) with the potential given by equation (\ref{DH-potential}). As an exact analytical solution is not available for this case, it is necessary to employ approximate methods. The potential of Debye-H\"{u}ckel has already been studied by different approaches \cite{Smith2, Rogers,Gonul}, in this article we use the variational method.

The trial function adopted for the ground state is constructed inspired on the successful function used for the unconfined Debye-H\"{u}ckel potential case \cite{Drigo2}. The specific function used for the ground state is given by:
\br
\Psi_{\mu} (r) = (r-R_c) e^{-(1-\frac{\mu}{2})r}(1 - e^{-\mu r}),
\er
where $R_c$ is the confining radius of the particles and $\mu$ is the variational parameter. The methodology adopted to introduce this trial function makes use of the algebraic relationship between the potential and the wave function, determined from Supersymmetric Quantum Mechanics \cite{Drigo4}. The term $(r - R_c)$ imposes confinement in the limit $r \rightarrow R_c$. In this limit, the potential tends to infinity and the wave function is equal to zero, configuring confinement by infinite walls. A similar approach has been applied to other potentials \cite{Drigo3, Silva2}.
	Following the variational method, the average energy value is calculated starting from the expression:
\br
\label{energylevels-confined}
E_v = \frac{\int_{0}^{R_c}{\Psi_{\mu}(r)^* \hat H \Psi_{\mu} dr}}{\int_{0}^{R_c}{\Psi_{\mu}(r)^*\Psi_{\mu}(r) dr}}
\er

The minimization of the variational energy, $E_v$, with respect to the variational parameter $\mu$, provides an upper bound for the true ground state energy. For a well-chosen trial function, the obtained energy value is a close approximation to the real eigenvalue. This study focuses exclusively on the ground state because any change in its eigenvalue is a sufficient indicator that quantum confinement effects must be considered for an accurate description of the system.

\section{Results}
The integrals in equation (\ref{energylevels-confined}) were calculated numerically (Mathematica software, Wolfram) and the minimum energy values were determined graphically as a function of the variational parameter $\mu$ for different confinement radii. The energy eigenvalues were calculated specifically for electrons  ($m_e = 1$ in atomic units) and the potential was consistently treated as attractive. 

Table 1 presents the resulting ground-state energies  ($E_e$)  for electrons subjected to the confined  Debye-H\"{u}ckel potential. This case clearly illustrates the drastic effect of spatial constraint on the electron cloud.  We observe a significant difference between the estimated eigenvalues for soft confinement (large confinement radii) and those calculated for the specific confinement radius determined from the viral capsid's internal pressure (approximately $8.7\, a_0$). As discussed in reference \cite{Drigo3}, our suggested trial function is expected to yield more accurate results for smaller confinement radii.  Nonetheless, the results consistently indicate that quantum effects, primarily those related to spatial confinement, are important for accurately describing the internal environment of viruses.

\begin{table}[H]
\caption{\label{tabone}Electron energy eigenvalues for the ground state $(E_e)$ of the Debye-H\"{u}ckel potential for different confinement radius values $(R_c)$. Values for variational parameters $(\mu)$ are also displayed. The values showed are in atomic units. } 
\begin{center}
\noindent \begin{tabular}{cccc} \hline
\multicolumn{1}{c} {$R_c (bohr)$ } & 
\multicolumn{1}{c} { $\mu (bohr^{-1})$ } & 
\multicolumn{1}{c} {$E_e (hartree)$ } & 
\\ \hline 
3.0& 3.385 & 0.637 \\  
4.0 &3.12	 & 0.375            \\    
5.0   &  2.956    & 0.2497         \\  
8.7  &  2.630   & 0.09303         \\ 
10.0  &  2.566    &  0.0728        \\   
20.0  &  2.328    &    0.02169      \\  
50.0 &  2.1556    &   $4.36 \,.10^{-3} $      \\
100.0&  2.08677    &   $ 1.28\,.10^{-3}$       \\ 
130.0&  2.06933    &   $ 8.03\,.10^{-4}$       \\
150.0&  2.06131    &   $ 6.23\,.10^{-4}$       \\
500.0&  2.02144    &   $ 7.23\,.10^{-5}$       \\\hline
\end{tabular}\\
\end{center}
\end{table}

\section{Conclusions} 
According to the proposed model, the effective confinement radius within the viral capsid is $R_c = 8.7 a_0$. Using this confinement radius, we determined the energy eigenvalues for an electron interacting via the Debye-H\"{u}ckel potential. With the parameters adopted by Forrey and Muthukumar \cite{Forrey} , the calculated energy eigenvalue is $E_e =  0.0930\,a.u.$, whereas for a weakly confined system $R_c = 500 \,a_0$ this value drops significantly to $E_e = 7.23 \,.10^{-5} \,a.u.$. These results, summarized in Table 1, clearly indicate that the ground-state electron energy for a system confined within a sphere of radius  $8.7 \,a_0$ is several orders of magnitude greater than that of an unconfined system.

The results strongly suggest that quantum effects are critical in describing the internal environment of viruses. This conclusion reinforces the observation made by Robson \cite{Robson}  , who noted that neglecting quantum contributions is a significant limitation in protein simulations. The specific limitation identified here is the disregard for quantum confinement effects in existing simulations. The internal environment of the viral capsid is subject to high pressure, which fundamentally alters the physical and chemical properties of the internal components.

Quantum effects are increasingly being introduced in the study of various aspects of biological macromolecules \cite{Marais, Cavasotto, Brookes}. The findings presented here indicate that the effect of quantum confinement also requires investigation. If this effect is ignored, computer simulations risk misinterpreting the processes that develop inside the virus. Consequently, the development of drugs or vaccines might be hindered by such a partial analysis of the macromolecules involved. 

Ultimately, this Supersymmetric Quantum Mechanics (SQM)-based framework represents a significant step towards a more complete and accurate model of viral behavior at the quantum level.

\end{document}